\documentclass[aps,prd,preprint,showpacs]{revtex4}
\usepackage{graphicx}
\usepackage{subfigure}
\usepackage{hyperref}
\usepackage{amsmath}
\begin{document}
\draft
\title{Anomalous Higgs Couplings at the LHeC  }

\author{A. Senol}\email{asenol@kastamonu.edu.tr}
\affiliation{Kastamonu University, Department of Physics, 37100,
Kastamonu, Turkey} \affiliation{Abant Izzet Baysal University,
Department of Physics, 14280, Bolu, Turkey}
\begin{abstract}
The discovery of Higgs boson plays a crucial role in understanding
the electroweak symmetry breaking sector. From now on, solving the
dynamics of this sector needs precision measurements of the
couplings of the Higgs boson to the standard model particles. In
this work, we investigate the constrains on the anomalous $HWW$ and
$HWW\gamma$ couplings, described by the dimension-six operators in
the effective Lagrangian, in a high energy envisaged ep collider
which is called Large Hadron electron Collider (LHeC). We obtained
the 95 \% confidence level limits on the couplings of anomalous
$HWW$ and $HWW\gamma$ vertex, with the design luminosity of 10
fb$^{-1}$ and electron beam energy of 140 GeV, through $ep\to \nu H
+X$, $\gamma p\to W H +X$ and $e\gamma\to W H \nu$ processes by
considering the new physics energy scale to be $\Lambda$= 1 TeV. The
sensitivity of the LHeC to the new physics scale is also briefly
discussed.
\end{abstract}
 \maketitle
\section{introduction}
After the discovery of a new boson being compatible with Standard
Model(SM) Higgs boson production and decay by ATLAS \cite{:2012gk}
and CMS \cite{:2012gu} Collaborations at the Large Hadron Collider
(LHC), the Electroweak Symmetry Breaking (EWSB) mechanism was
verified experimentally leading to open up a gateway for new
research field in particle physics. Now, the constraints on
couplings of Higgs boson with the SM particles need to be
reconsidered due to the fact that the precision measurements of its
couplings will give us detailed information on EWSB of the SM and
beyond. Therefore, we focus on anomalous couplings of $HWW$ and
$HWW\gamma$ vertex in ep collision where some advantages over the
LHC for precision measurements such as: the ability to separate
backward scattering and forward scattering due to characteristic ep
kinematics and an anomalous $HWW$ vertex will be free from possible
contaminations of other Higgs boson-electroweak vector boson
couplings.

Recently, there has been a new ep collider project, the Large Hadron
Electron Collider (LHeC) \cite{AbelleiraFernandez:2012cc}, in which
a newly built electron beam of 60 GeV, to possibly 140 GeV, energy
collides with the intense hadron beams of the LHC (7 TeV) and with
the design luminosity of $10^{33}$ cm$^{-2}$ s$^{-1}$. The physics
programme is purposed to a search of the energy frontier,
complementing the LHC and its discovery potential for physics beyond
the Standard Model.

There have been several studies for anomalous couplings of $HWW$
vertex in the literature which focus on future linear $e^+ e^-$
collider
\cite{Hagiwara:1993ck,Hagiwara:1993qt,Gounaris:1995mx,Kilian:1996wu,Lietti:1996gd,Biswal:2008tg,Biswal:2005fh}
and its $e\gamma$ \cite{Choudhury:2006xe,Sahin:2008jc}  and
$\gamma\gamma$
\cite{Gounaris:1995gm,Banin:1998ap,Han:2005pu,Sahin:2008qp} modes ,
hadron colliders \cite{deCampos:1997ez,GonzalezGarcia:1999fq,
He:2002qi, Zhang:2003it,
Hankele:2006ma,Kanemura:2008ub,Desai:2011yj,Bonnet:2011yx,Corbett:2012dm}
and also ep collider \cite{Biswal:2012mp}. In Ref.
\cite{Biswal:2012mp}, the constrains on anomalous CP-conserving and
CP-violating couplings of $HWW$ vertex coming from dimension-five
operators in the effective Lagrangian are studied. Furthermore, we
will analyze the anomalous couplings of $HWW$ and $HWW\gamma$ vertex
coming from dimension-six operators in the effective Lagrangian.

The Higgs-vector boson vertices are uniquely assigned in the SM. In
some models deviations from these vertices appear, such as
non-pointlike character of boson and through interactions beyond the
SM. We do not have a specific model to analyze for the effect of
non-SM couplings. We investigated anomalous Higgs-vector boson
couplings in a model independent way by means of effective
non-renormalizable Lagrangian approach which keep the SM gauge group
\cite{Buchmuller:1985jz, Hagiwara:1993qt}

\begin{equation}
{\cal L}_{eff} = {\cal L}_{SM} +
\sum_{k=1}^{\infty}\frac{1}{(\Lambda^2)^k} \sum_i f_i^{(k)}Q_i^{d_k}
\quad , \label{eq:lagrangian}
\end{equation}
where $d_k = 2 k + 4$ denotes the dimension of operators and
$\Lambda$ is the energy scale of new interactions. We study only to
complete set of the dimension-6 operators.

In this framework, there are only two relevant operators that Higgs
boson couplings to electroweak vector bosons:
\begin{equation}
\frac{1}{\Lambda^2}\left\{\frac{1}{2}f_{\varphi}\partial_{\mu}
(\Phi^+\Phi)\partial^{\mu}(\Phi^+\Phi) + f_{WW}\Phi^+({\hat
W}_{\mu\nu}\hat W^{\mu\nu})\Phi\right\}. \label{eq:1overlambda}
\end{equation}

We use the formalism of \cite{Boos:1996ud} in writing for $HWW$ and
$HWW\gamma$ vertices in unitary gauge which follow from the
effective Lagrangian (\ref{eq:lagrangian}) and
(\ref{eq:1overlambda}):

\begin{eqnarray}
\Gamma^{HWW}_{\mu \nu}(p,q,r) = \frac{e M_W}{s_W}\biggl\{(1 -
\frac{1}{4}f_{\varphi}\frac{v^2}{\Lambda^2})g_{\mu\nu} +
f_{WW}\frac{1}{\Lambda^2}[g_{\mu\nu}(q\cdot r) -
q_{\nu}r_{\mu})]\biggr\} \label{HWW}
\end{eqnarray}
 and
\begin{eqnarray}
\Gamma^{HWW\gamma}_{\mu \nu\alpha}(p,q,r, l) = \frac{e^2 M_W}{s_W} 2
f_{WW} \frac{1}{\Lambda^2}  \biggl\{  g_{\mu\nu}(q - r)_{\alpha} -
q_{\nu}g_{\mu\alpha} + r_{\mu}g_{\nu\alpha}\biggr\}\label{HWWg}
\end{eqnarray}
where $v = \frac{2 M_W}{e}s_W$ is the vacuum expectation value; $p,
q, r$ and $l$ are the momenta of the $H, W^+, W^-$ and $\gamma$
fields, respectively. $\mu, \nu$ and $ \alpha$ denote the $W$'s and
$ \gamma$ fields, respectively. If the values of $f_{WW}$ and
$f_{\varphi}$ are zero in $\Gamma^{HWW}_{\mu \nu}$ vertex, it
corresponds to the SM vertex at tree level. The second vertex
$\Gamma^{HWW\gamma}_{\mu \nu\alpha}$ does not occur in the SM at
tree level. All calculations were performed by means of computer
package the CalcHEP \cite{Belyaev:2012qa}, after implementation of
the vertices (\ref{HWW}) and (\ref{HWWg}) with taking $\Lambda=1$
TeV and $m_H$= 125 GeV.
\section{ The cross sections of  $ep\to \nu H +X$, $\gamma p\to W H +X$ and $e\gamma\to W H
\nu$ Processes}
\begin{figure*}[htbp!]
  \includegraphics[width=4cm]{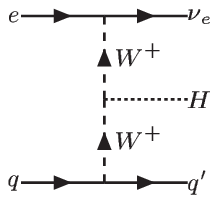}\\
 \caption{Tree-level Feynman diagram for the process $ep\to \nu H +X$.}\label{fig1}
\end{figure*}

\begin{figure*}[htbp!]
  \includegraphics[width=8cm]{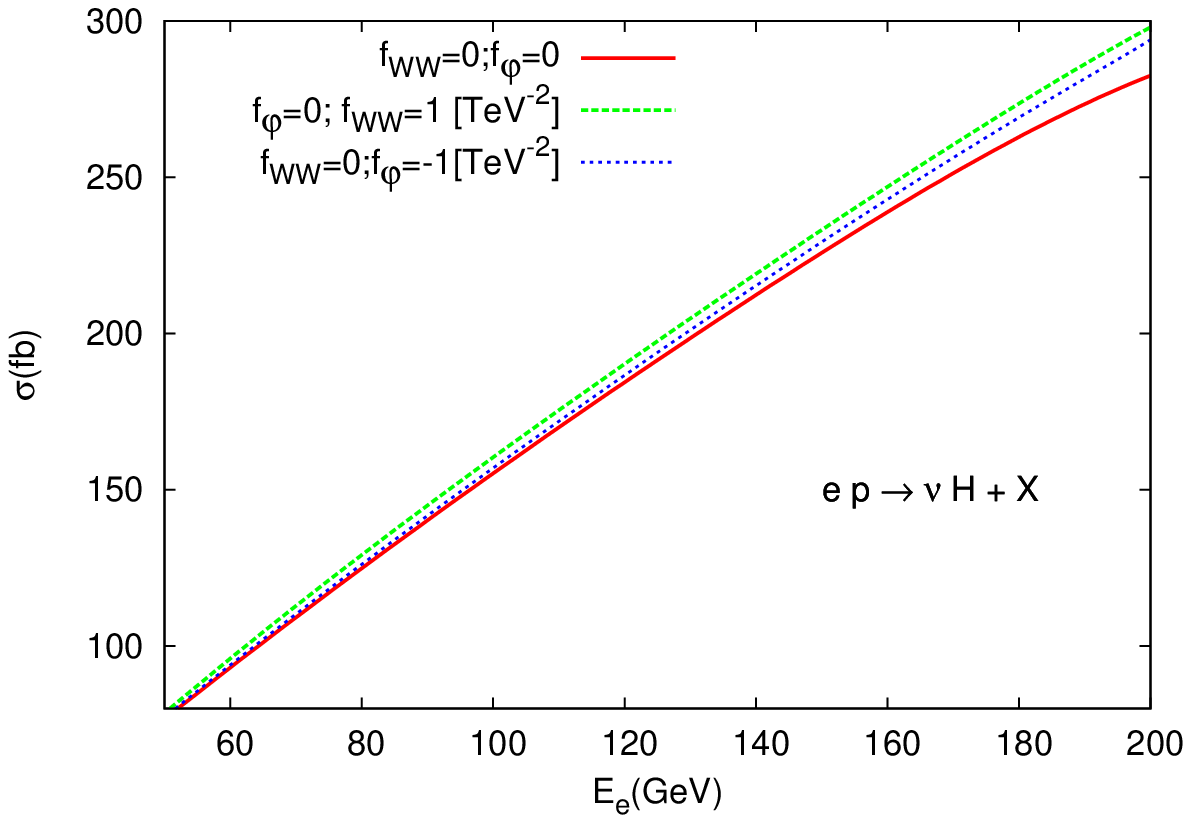} \includegraphics[width=8cm]{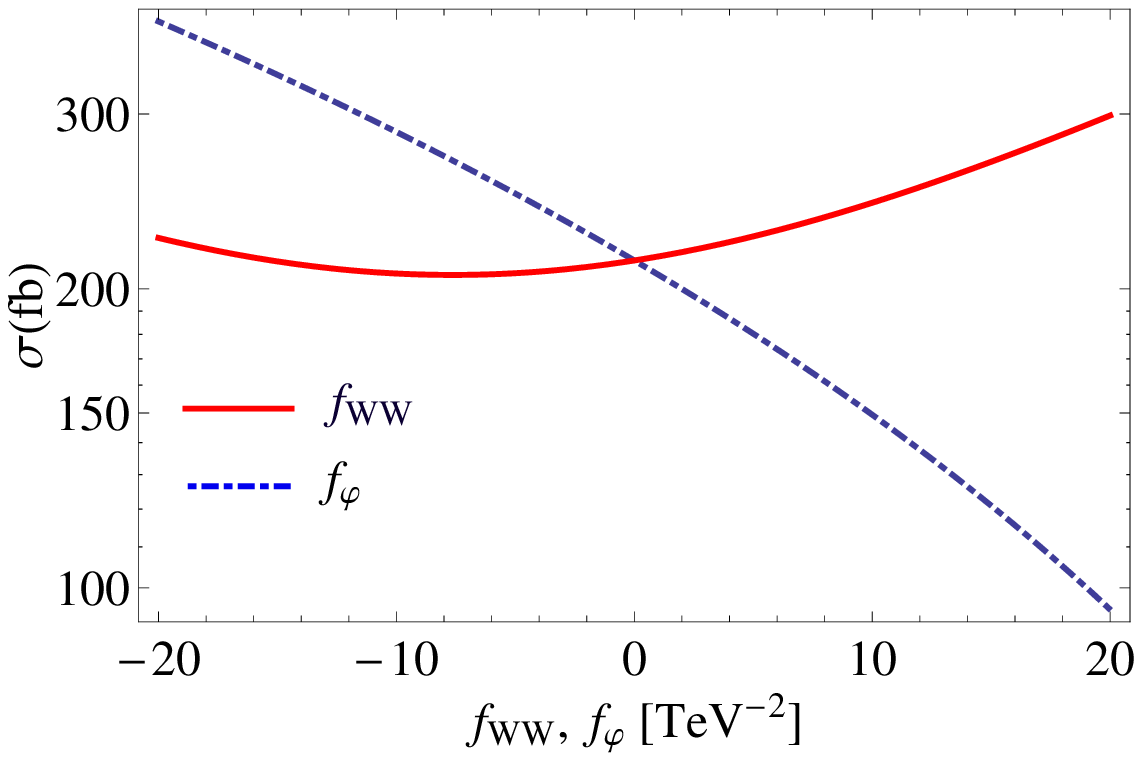}\\
  \caption{The total cross sections for the process $ep\to \nu H+X$
  including only anomalous $HWW$ couplings in ep collisions at the LHeC. The plot on the left displays incoming electron energy dependence with taking $m_H$= 125 GeV.
  The plot on the right shows dependence on anomalous couplings $f_{WW}$ (solid line) and $f_{\varphi}$ (dashed line) with taking $E_e$=140 GeV and $m_H$= 125 GeV.}
\label{fig5}
\end{figure*}

The production mechanism for a Higgs boson in the
$WW$ fusion at the LHeC is $ep\to \nu H +X$ as shown in
Fig.~\ref{fig1}. This process has a single Feynman diagram involving
the $HWW$ vertex.
In the left panel of Fig. \ref{fig5}, we display the total cross
sections depending on incoming electron energy for the reaction
$ep\to \nu H+X$ including only anomalous $HWW$ coupling with taking
$f_{WW}$($f_{\varphi}$)=1 (0) TeV$^{-2}$, $f_{\varphi}(f_{WW})$=-1
(0) TeV$^{-2}$ and $f_{WW}$=$f_{\varphi}=0$ for illustration purpose
by using parton distribution functions library CTEQ6L
\cite{Pumplin:2002vw}. The calculated total cross sections of the
$ep\to \nu H +X$ process with taking 140 GeV of energy of incoming
electron as function of the anomalous couplings $f_{WW}$ and
$f_{\varphi}$ is shown in the right panel of Fig. \ref{fig5}. Here
and henceforth, only one of the coupling parameter is kept from
zero. From left panel of Fig. \ref{fig5} we can see that, the only
contribution from the SM part of the Eq. (\ref{HWW}) in the case of
$f_{WW}$=$f_{\varphi}=0$ (solid line), from both $f_{\varphi}$
($f_{WW}$) coupling and SM part in the case of $f_{\varphi}=-1 (0)$
TeV$^{-2}$, $f_{WW}$=0 (1) TeV$^{-2}$ to $ep\to \nu H +X$ process.
Here, we see much larger deviation from SM cross sections on the
positive region of the anomalous coupling $f_{\varphi}$ due to the
negative factor front the anomalous coupling $f_{\varphi}$ in the
Eq. (\ref{HWW}).
\begin{figure*}[htbp!]
  \includegraphics[width=14cm]{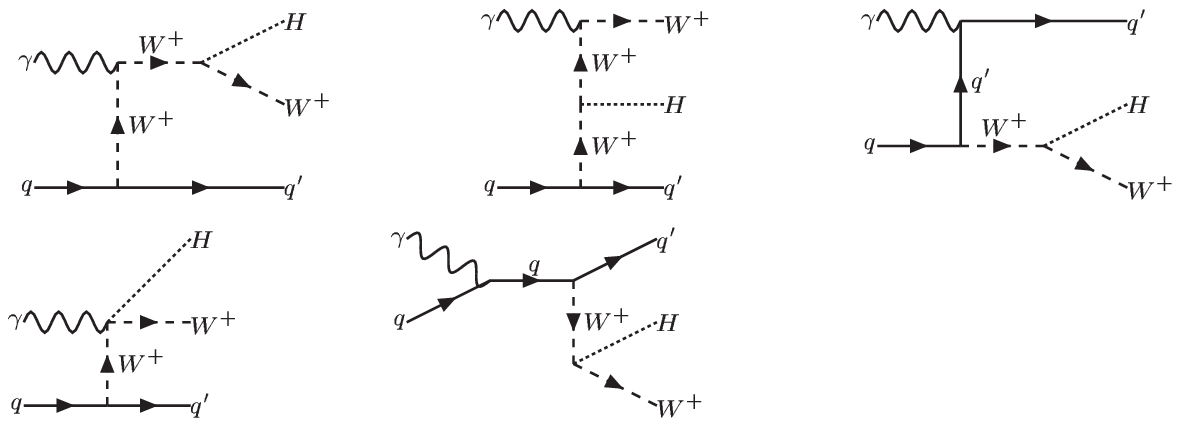}\\
 \caption{Tree-level Feynman diagrams for the process $\gamma p\to W H +X$.}\label{fig2}
\end{figure*}

\begin{figure*}[htbp!]
\includegraphics[width=8cm]{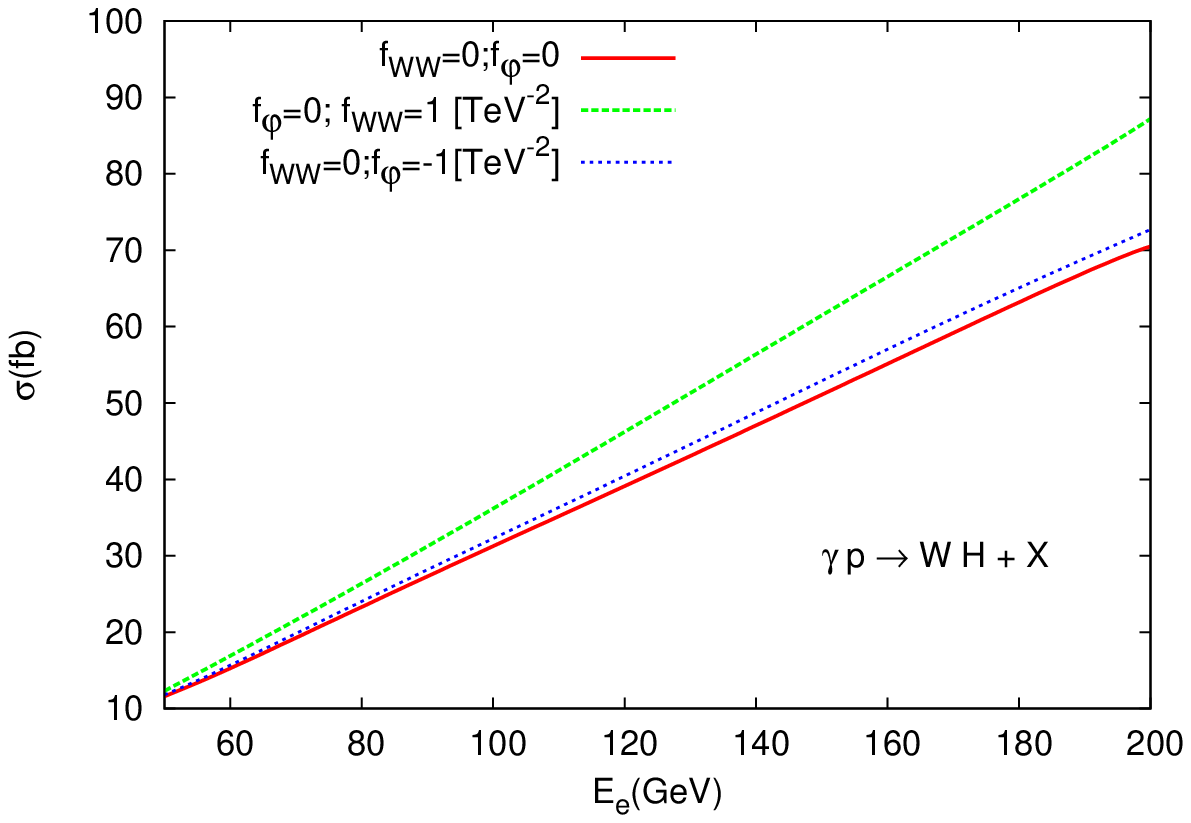}\includegraphics[width=8cm]{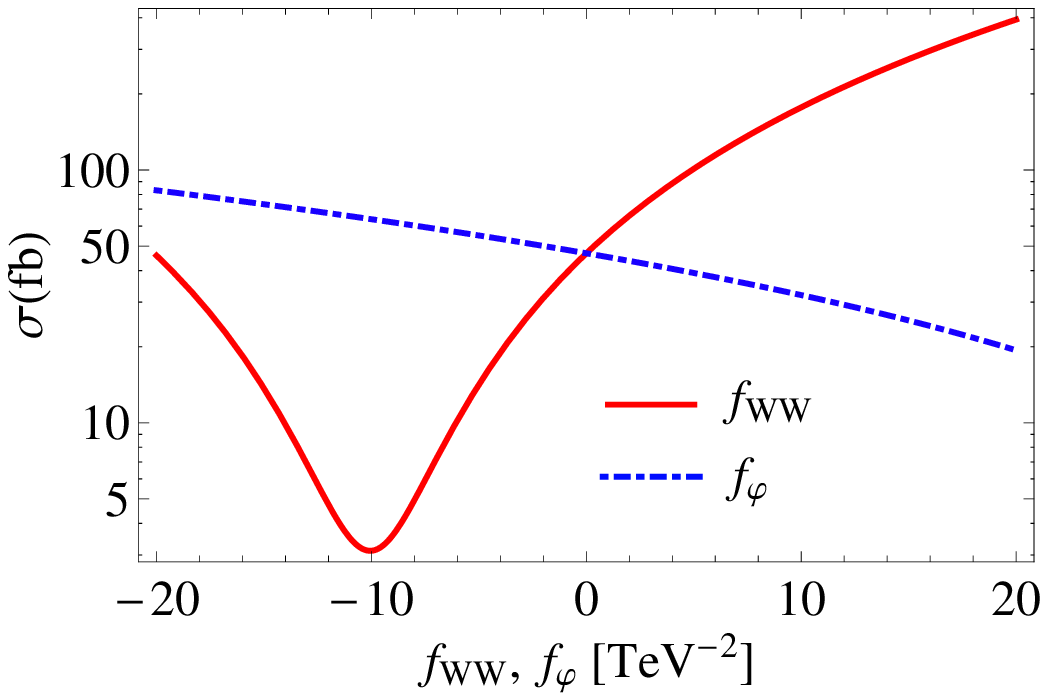}\\
\caption{The total cross sections depending on incoming electron
energy for $\gamma p\to W H +X$ including anomalous $HWW$ and
$HWW\gamma$ couplings in ep collisions at the LHeC.The plot on the
left displays incoming electron energy dependence with taking $m_H$=
125 GeV. The plot on the right shows dependence on anomalous
couplings $f_{WW}$ (solid line) and $f_{\varphi}$ (dashed line) with
taking $E_e$=140 GeV and $m_H$= 125 GeV.}\label{fig6}
\end{figure*}

Efficient $\gamma p$ collisions can be realized with real $\gamma$,
produced using Compton back scattering of laser beam off the high
energy electron beam, only on the base of linac ring type $ep$
colliders \cite{Kaya:2012tn}. In this framework, we consider $\gamma
p\to W H +X$ reaction to see the effect of both $HWW$ and
$HWW\gamma$ couplings. The tree-level diagrams of the process
$\gamma p\to W H +X$ are depicted in Fig. \ref{fig2}. We present the
total cross section as function of incoming electron beam energy for
this process by using the spectrum of photons scattered backward
from the interaction of laser light with the high energy electron
beam \cite{Ginzburg:1981vm} in case $f_{WW}$=$f_{\varphi}=0$,
$f_{WW}(f_{\varphi})=1 (0)$ TeV$^{-2}$ and $f_{\varphi}(f_{WW})$=-1
(0) TeV$^{-2}$ in the left panel of Fig. \ref{fig6}. As we can see,
contribution of the $HWW\gamma$ vertex, described in Eq.
(\ref{HWWg}), leads to an increase of about two orders in the cross
section. Total cross sections of $\gamma p\to W H +X$ process as a
functions of $f_{WW}$ and $f_{\varphi}$ is shown in the left panel
of Fig. \ref{fig6}.

\begin{figure*}[htbp!]
  \includegraphics[width=10cm]{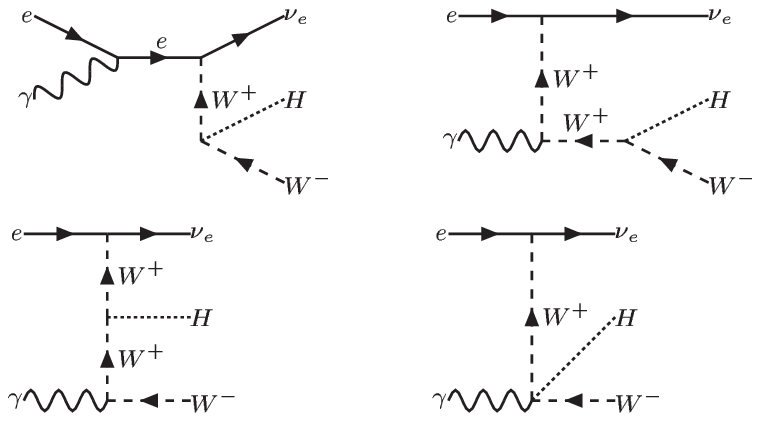}\\
 \caption{Tree-level Feynman diagrams for the process $e\gamma\to W H \nu$.}\label{fig3}
\end{figure*}

\begin{figure*}[htbp!]
  \includegraphics[width=8cm]{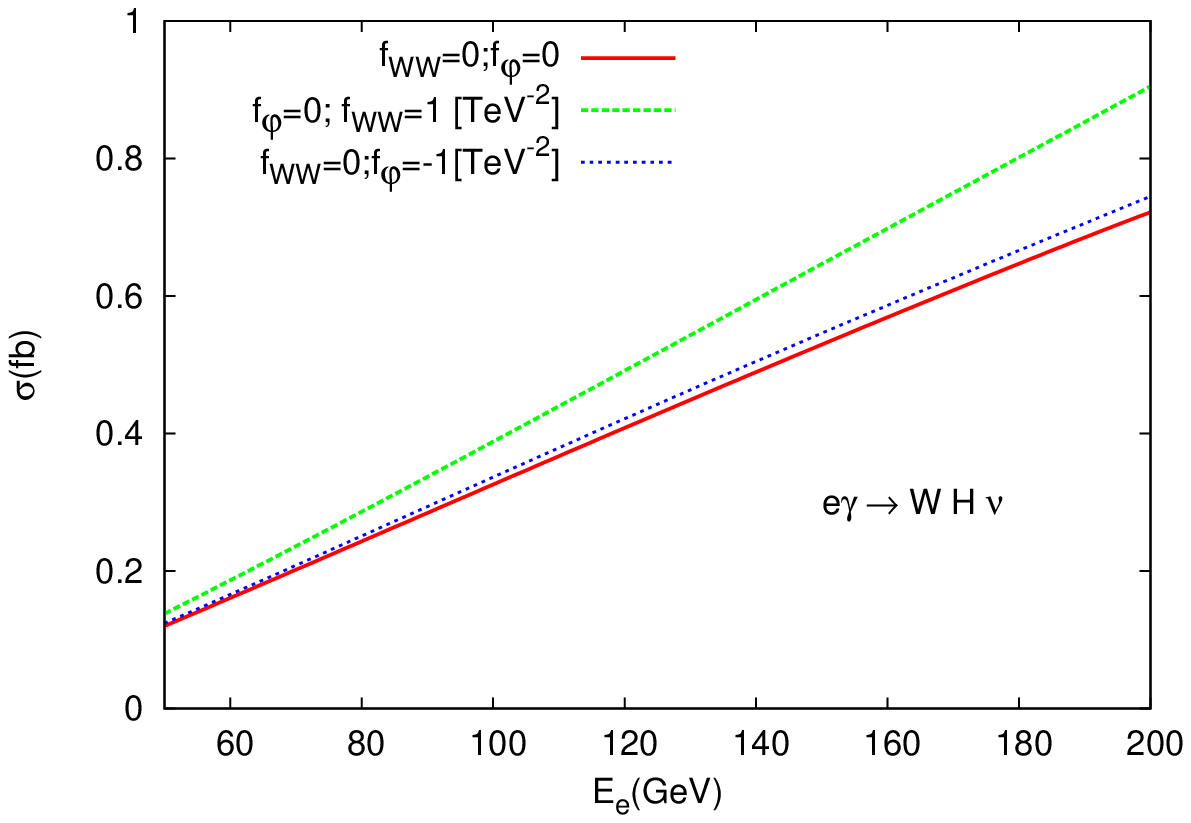} \includegraphics[width=8cm]{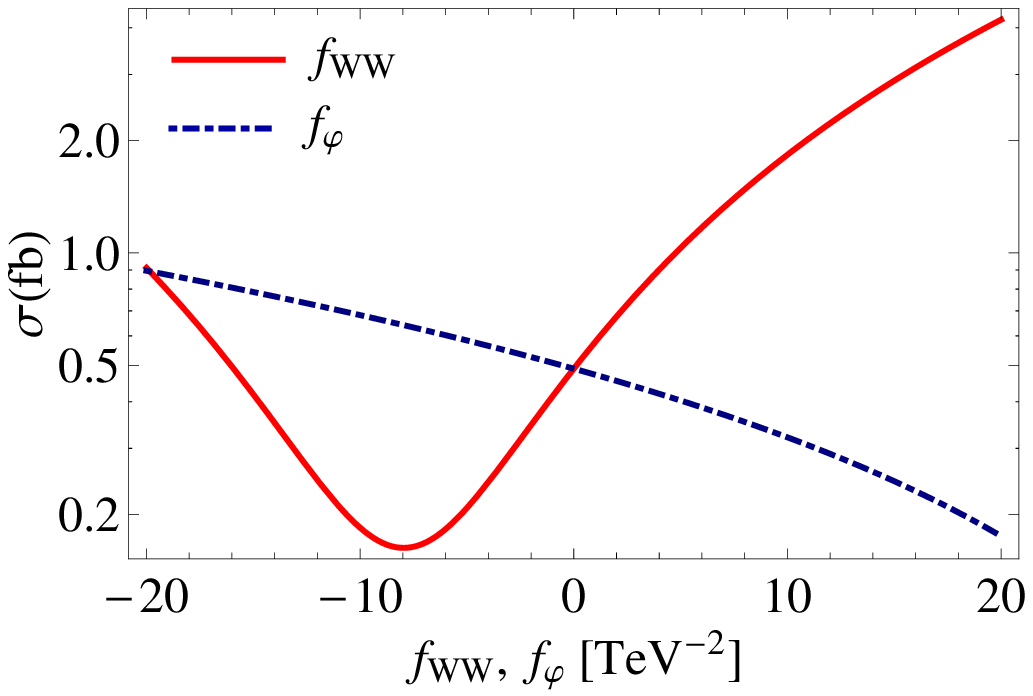}\\
 \caption{The total cross sections depending on incoming electron
energy for $e\gamma\to WH\nu$ including anomalous $HWW$ and
$HWW\gamma$ couplings in ep collisions at the LHeC.The plot on the
left displays incoming electron energy dependence with taking $m_H$=
125 GeV. The plot on the right shows dependence on anomalous
couplings $f_{WW}$ (solid line) and $f_{\varphi}$ (dashed line) with
taking $E_e$=140 GeV and $m_H$= 125 GeV. }\label{fig4}
\end{figure*}

The another mode of ep colliders is $e\gamma $ option where $\gamma$
is elastic photon emission coming from proton. The equivalent photon
spectrum is described by the equivalent photon approximation (EPA)
\cite{Budnev:1974de} which embedded in CalcHEP. The $e\gamma\to W H
\nu$ process in ep collision is described by tree-level diagrams in
Fig. \ref{fig3}. These diagrams contain anomalous $HWW$ and
$HWW\gamma$ couplings. In Fig. \ref{fig4}, we plot the total cross
section depending on incoming electron energy for
$f_{WW}$=$f_{\varphi}=0$, $f_{WW}(f_{\varphi})=1 (0)$ TeV$^{-2}$ and
$f_{\varphi}(f_{WW})$=-1 (0) TeV$^{-2}$ (left panel) and as
functions of $f_{WW}$ and $f_{\varphi}$ with taking $E_e$=140 GeV
(right panel) by using EPA.

\section{Limits on the anomalous Higgs couplings}
One-parameter $\chi^2$ test was applied without a systematic error
to obtain 95\% confidence level (C.L.) on the upper limits of the
$f_{\varphi}$ and $f_{WW}$. The $\chi^2$ function is
\begin{eqnarray}
\chi^{2}=\left(\frac{\sigma_{SM}-\sigma(f_{\varphi},
f_{WW})}{\sigma_{SM} \,\, \delta}\right)^{2}
\end{eqnarray}
where $\delta=\frac{1}{\sqrt{N}}$ is the statistical error. The
number of events are given by $N=\sigma_{SM}L_{int}$ where $L_{int}$
is the integrated luminosity. When calculating number of events we
assume all $W$ bosons decay leptonically in the final state, the
dominant Higgs boson decay to $b\bar b$, the efficiency for
b-tagging to be $\epsilon=60\%$ and the fake rejection factors of
0.01 for light quarks. And also we applied cuts for missing
transverse energy (MET) for neutrinos to be MET $>25$ GeV,
transverse momentum of quarks to be $p_T^{b,j}> 30$ GeV and
pseudorapidity of quarks to be $|\eta|^{b,j}< 2.5$. With assuming
these restrictions, we have calculated total cross sections
$\sigma_{SM}=0.047$ pb for $ep\to \nu H+X$, $\sigma_{SM}=7.61\times
10^{-3}$ pb for $\gamma p\to H W + X$ and $\sigma_{SM}=3.72\times
10^{-4}$ pb for $e \gamma \to H W \nu$ processes.

\begin{figure*}[htbp!]  
  \includegraphics[width=8cm]{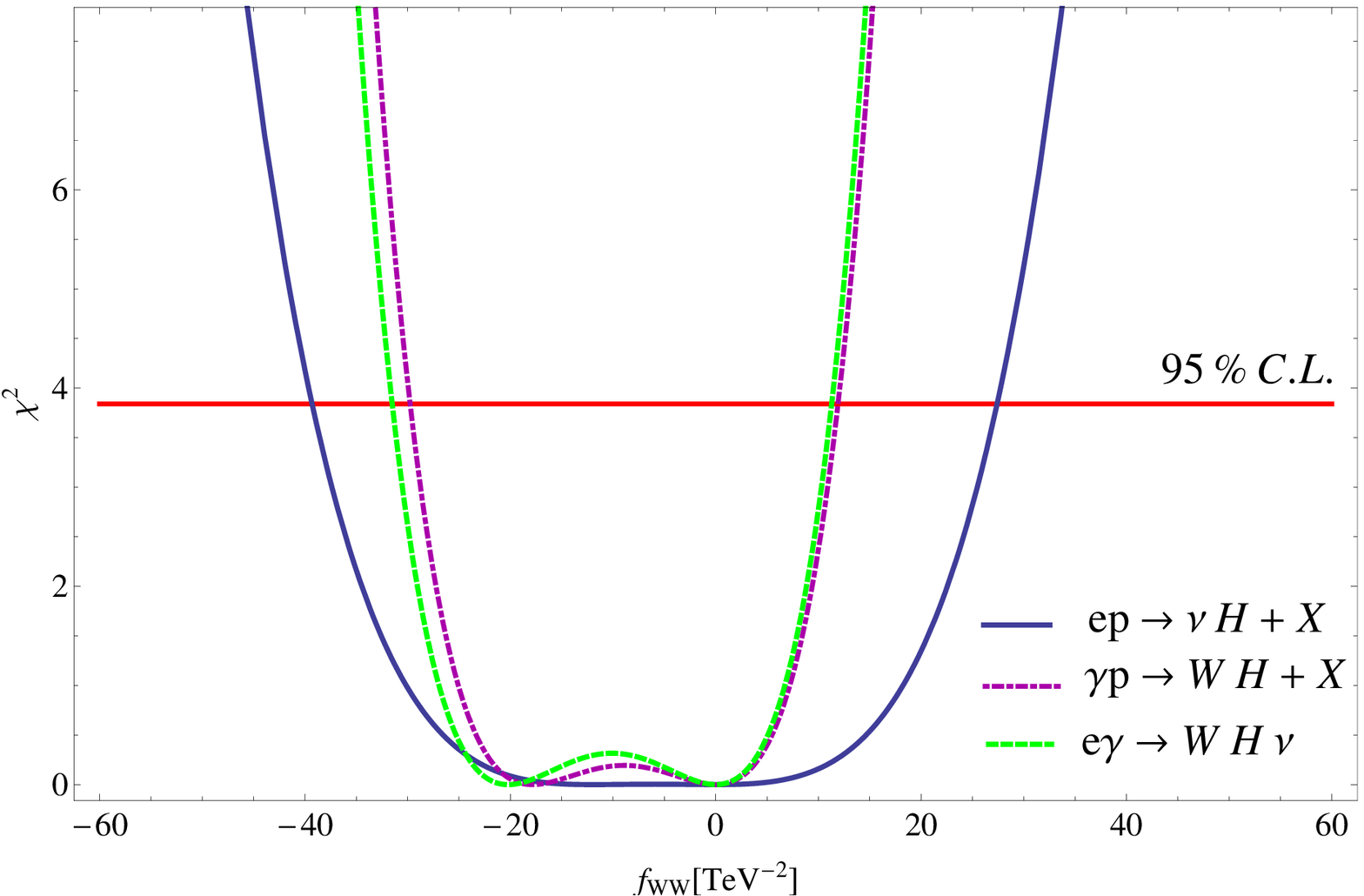}\includegraphics[width=8cm]{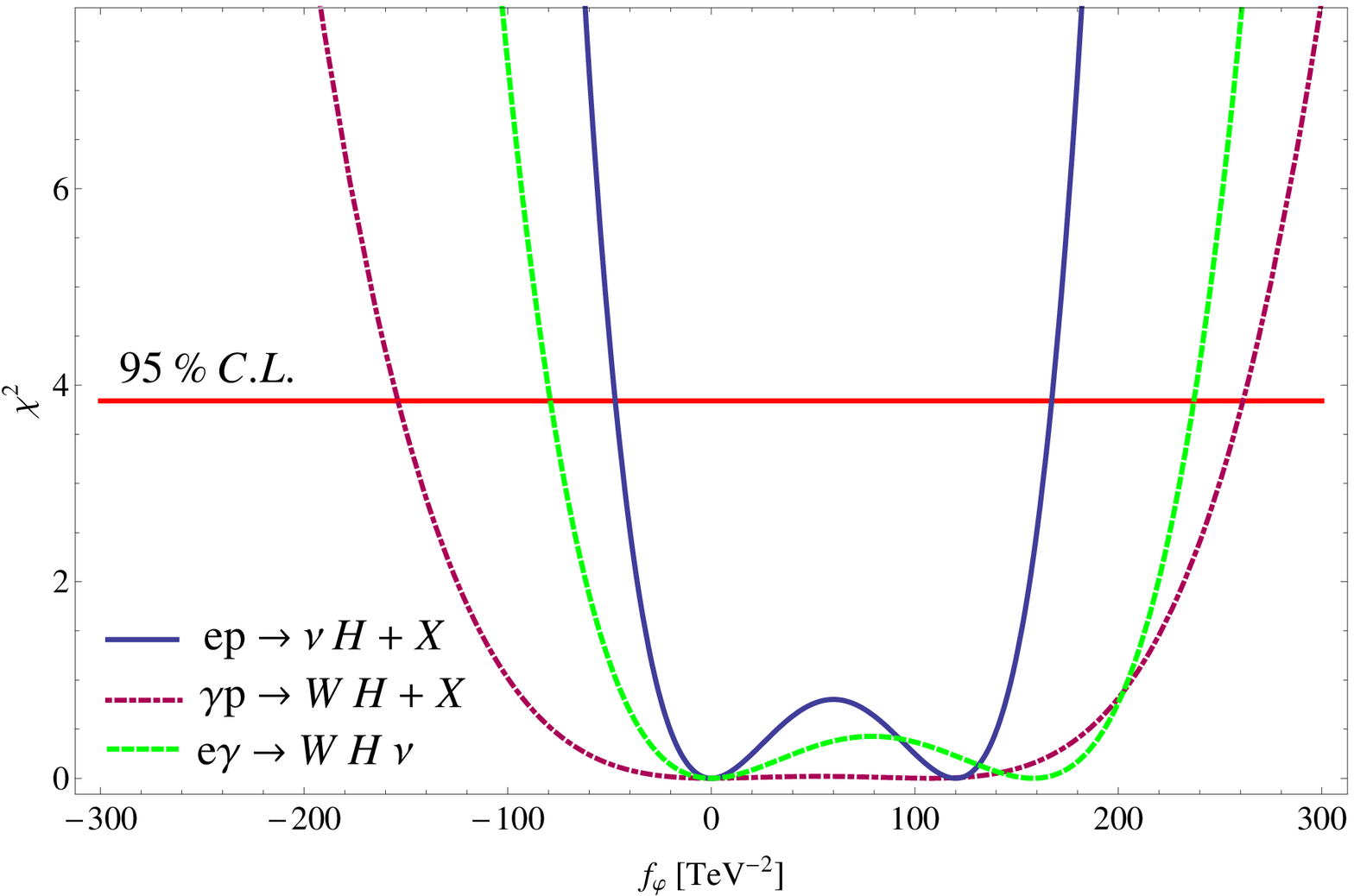}\\
 \caption{$\chi^2$ as a function of $f_{WW}$
(left panel) and $f_{\varphi}$ (right panel) through $ep\to \nu H
+X$, $\gamma p\to W H +X$ and $e\gamma\to W H \nu $ with $E_e$=140
GeV and design luminosity of 10 fb$^{-1}$.}\label{fig11}
\end{figure*}

In Fig. \ref{fig11}, we exhibited $\chi^2$ as a function of $f_{WW}$
(left panel) and $f_{\varphi}$ (right panel) through $ep\to \nu H
+X$, $\gamma p\to W H +X$ and $e\gamma\to W H \nu $ with $\Lambda=1$
TeV, $E_e$=140 GeV and design luminosity, $L=10$ fb$^{-1}$. A
distinct feature of this figure is that the limiting on anomalous
couplings to see clearly at 95\% C.L.. If the LHeC has collected 10
fb$^{-1}$ of data, the bounds on $f_{WW}$ would be (-39.3, 27.4)
TeV$^{-2}$ for $ep\to \nu H +X$, (-29.8, 11.9) TeV$^{-2}$ for
$\gamma p\to W H +X$, (-31.5, 11.3) TeV$^{-2}$ for $e\gamma\to W H
\nu$ process and $f_{\varphi}$ would be (-47.2, 167.2)  TeV$^{-2}$
for $ep\to \nu H +X$, (-153.9, 261.2) TeV$^{-2}$ for $\gamma p\to W
H +X$, (-79.0, 237.1) TeV$^{-2}$ for $e\gamma\to W H \nu$ at 95\%
C.L. While the indirect 95\% C.L. constraints of the L3
collaboration \cite{Achard:2004kn,Hankele:2006ma} for $f_{WW}$ are
in the interval of (-26.84, 26.84) TeV$^{-2}$  with taking $m_H$=120
GeV and (-7.0, 10) TeV$^{-2}$  form available data from Tevatron and
LHC at 90\% C.L. \cite{Corbett:2012dm,Corbett:2012ja}.

In Table~\ref{1D}, we give 95\% C.L. bounds of the couplings
$f_{WW}$ and $f_{\varphi}$ for three different values of new physics
energy scale, $\Lambda$, at $ep\to \nu H +X$, $\gamma p\to W H +X$
and $e\gamma\to W H \nu$ processes with the design luminosity of 10
fb$^{-1}$ and electron beam energy of 140 GeV. The fact that the
sensitivity of coupling $f_{\varphi}$ is more rapidly decrease,
compared to $f_{WW}$, when the scale of new physics increase. On the
other hand, we can see a faster drop on the sensitivity of coupling
$f_{WW}$ in $ep\to \nu H +X$ process, compared to $e\gamma\to W H
\nu$, at $\Lambda$=3 TeV.

\begin{table}
\caption{Variations of $f_{WW}$ and $f_{\varphi}$ couplings with
respect to $\Lambda$ at 95\% C.L. for  $ep\to \nu H +X$, $\gamma
p\to W H +X$ and $e\gamma\to W H \nu $ processes with $E_e$=140 GeV
and design luminosity of 10 fb$^{-1}$.\label{1D}}
\begin{tabular}{lcccccccc}
  \hline\hline
  &\multicolumn{8}{l}{~~~~~~~~~~~$ep\to \nu H +X$~~~~~~~~~~~~~~~~~~~~~$\gamma p\to W H +X$~~~~~~~~~~~~~~~~~~~~~~ $e\gamma\to W H \nu $} \\ [0.5ex]
$\Lambda$ (TeV) &$f_{WW}$ &$f_{\varphi}$    &&$f_{WW}$ &$f_{\varphi}$ && $f_{WW}$&$f_{\varphi}$
\\\hline
  1             & (-39.3, 27.4)&(-47.2, 167.2)& & (-29.8, 11.9)&(-153.9, 261.2) && (-31.5, 11.3)&(-79.0, 237.1) \\
  2             & (-44.7, 37.2)&(-175.7, 575.3) && (-117.2, 48.3)&(-495.1, 789.2)& & (-129.9, 43.8)&(-306.9, 968.9) \\
  3             & (-338.3, 218.3)&(-281.7, 973.8) && (-238.5, 115.9)&(-714.6, 1004.3)&& (-288.4, 99.7)&(-634.5, 2395.6) \\
  \hline\hline
\end{tabular}
\end{table}
\section{conclusion}
In this work, we focused on couplings of $HWW$ and $HWW\gamma$
vertices to constrain deviations from the SM behavior leading the
effects of dimension-six effective operators by considering the new
physics energy scale to be $\Lambda$= 1 TeV. We have examined these
effects at $ep\to \nu H +X$, $\gamma p\to W H +X$ and $e\gamma\to W
H \nu$ processes at the LHeC to compare which can give the best
limits on the anomalous couplings. Best limits on $f_{WW}$ are
obtained about (-39.3, 27.4) TeV$^{-2}$ at $ep\to \nu H +X$ process,
(-29.8, 11.9) TeV$^{-2}$ at $\gamma p\to W H +X$ and (-31.5, 11.3)
TeV$^{-2}$ at $e\gamma\to W H \nu$ process and the limits on
$f_{\varphi}$ are obtained about (-42.45, 7.18) TeV$^{-2}$ at $ep\to
\nu H +X$,(-153.9, 261.2) TeV$^{-2}$ at $\gamma p\to W H +X$,
(-79.0, 237.1) TeV$^{-2}$ in $e\gamma\to W H \nu$ at 95 \% C.L. with
the design luminosity value. The sensitivity on anomalous couplings,
$f_{WW}$ and $f_{\varphi}$ , with respect to new physics scale are
investigated. It is shown that the sensitivity of $f_{\varphi}$
rapidly decrease, compared to $f_{WW}$, when the scale of new
physics increase. We cannot simply compare our results on $f_{WW}$
and $f_{\varphi}$ one to one with the experimental limits obtained
by various sources due to the different conventions adopted in the
literature. However, the current experimental limits are of same
order as our bounds. As well as, an integrated luminosity of 10
$fb^{-1}$ would be enough to probe small values of anomalous Higgs
couplings. Nevertheless, the LHeC is a suitable platform to
complement the LHC results for searching of anomalous $HWW$ and
$HWW\gamma$ couplings in $ep\to \nu H +X$ process as well as $\gamma
p\to W H +X$ and $e\gamma\to W H \nu $ processes.
  
\end{document}